\documentclass[10pt,twocolumn,letterpaper]{article}

\usepackage{cvpr}
\usepackage{times}
\usepackage{epsfig}
\usepackage{graphicx}
\usepackage{amsmath}
\usepackage{amssymb}

\usepackage[breaklinks=true,bookmarks=false]{hyperref}

\cvprfinalcopy 

\def\cvprPaperID{5238} 

\begin{document}

\title{X2CT-GAN: Reconstructing CT from Biplanar X-Rays with Generative Adversarial Networks}

\author{
Xingde Ying\footnotemark[1]\ \ , Heng Guo\footnotemark[1]\ \ , Kai Ma, Jian Wu, Zhengxin Weng, and Yefeng Zheng\\
}

\maketitle
\renewcommand{\thefootnote}{\fnsymbol{footnote}}
\footnotetext[1]{Equally-contributed.}
\renewcommand{\thefootnote}{\arabic{footnote}}
\begin{abstract}
Computed tomography (CT) can provide a 3D view of the patient's internal organs, facilitating disease diagnosis, but it incurs more radiation dose to a patient and a CT scanner is much more cost prohibitive than an X-ray machine too. 
Traditional CT reconstruction methods require hundreds of X-ray projections through a full rotational scan of the body, which cannot be performed on a typical X-ray machine.
In this work, we propose to reconstruct CT from two orthogonal X-rays using the generative adversarial network (GAN) framework.
A specially designed generator network is exploited to increase data dimension from 2D (X-rays) to 3D (CT), which is not addressed in previous research of GAN.
A novel feature fusion method is proposed to combine information from two X-rays.
The mean squared error (MSE) loss and adversarial loss are combined to train the generator, resulting in a high-quality CT volume both visually and quantitatively.
Extensive experiments on a publicly available chest CT dataset demonstrate the effectiveness of the proposed method.
It could be a nice enhancement of a low-cost X-ray machine to provide physicians a CT-like 3D volume in several niche applications.
\end{abstract}

\section{Introduction}
Immediately after its discovery by Wilhelm Röntgen in 1895, X-ray found wide applications in clinical practice. It is the first imaging modality enabling us to non-invasively see through a human body and diagnose changes of internal anatomies. However, all tissues are projected onto a 2D image, overlaying each other. While bones are clearly visible, soft tissues are often difficult to discern.
Computed tomography (CT) is an imaging modality that reconstructs a 3D volume from a set of X-rays (usually, at least 100 images) captured in a full rotation of the X-ray apparatus around the body.
One prominent advantage of CT is that tissues are presented in the 3D space, which completely solves the overlaying issue. However, a CT scan incurs far more radiation dose to a patient (depending on the number of X-rays acquired for CT reconstruction). Moreover, a CT scanner is often much more cost prohibitive than an X-ray machine, making its less accessible in developing countries \cite{world2011baseline}. 

\begin{figure}
 	\centering
	\includegraphics[width=1.0\linewidth]{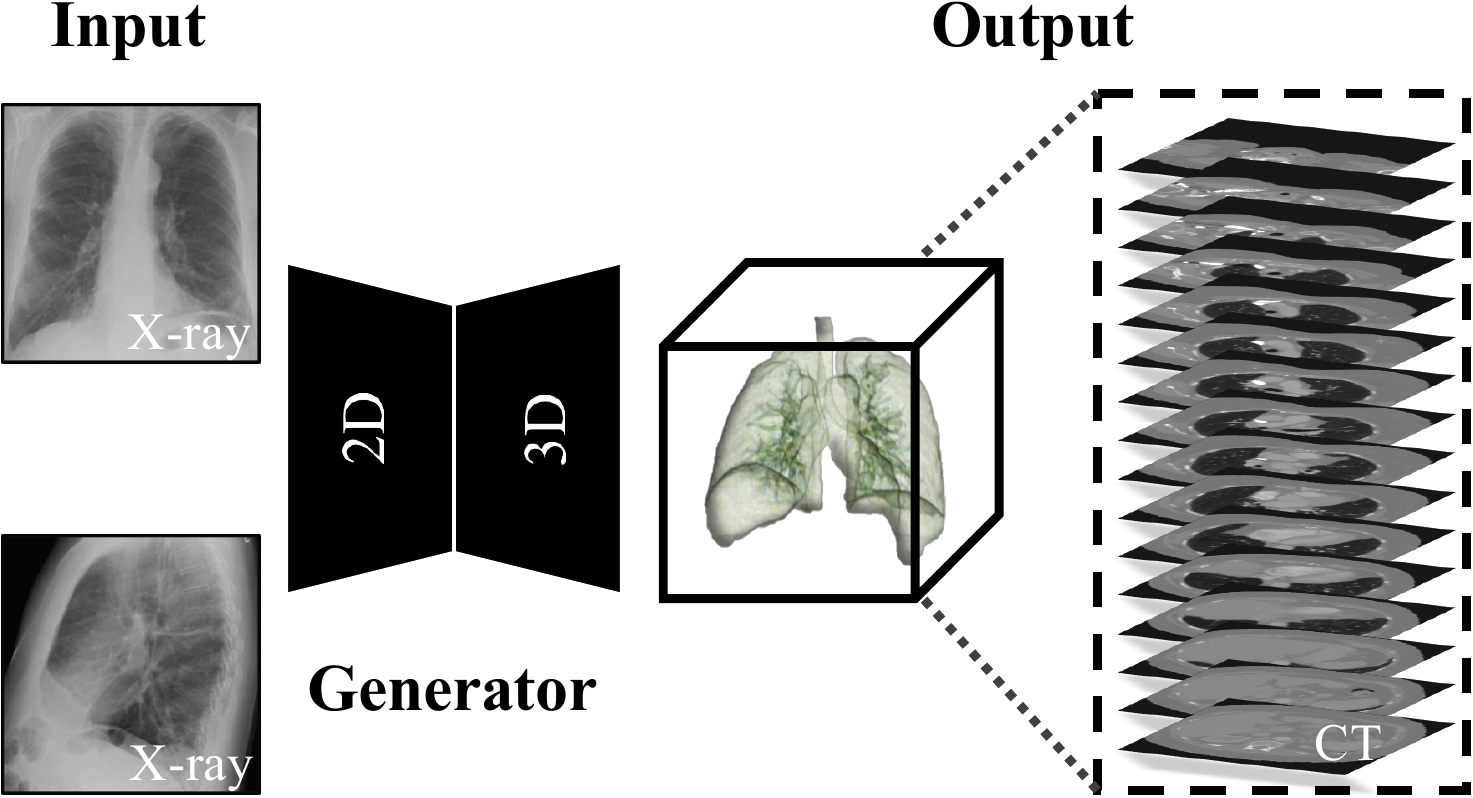}
	\caption{Illustration of the proposed method. The network takes 2D biplanar X-rays as input and outputs a 3D CT volume.}
	\label{motivation}
\end{figure}

With hundreds of X-ray projections, standard reconstruction algorithms, \eg, filtered back projection or iterative reconstruction, can accurately reconstruct a CT volume \cite{Herman2009CT}. However, the data acquisition requires a fast rotation of the X-ray apparatus around the patient, which cannot be performed on a typical X-ray machine. In this work, we propose to reconstruct a CT volume from biplanar X-rays that are captured from two orthogonal view planes. The major challenge is that the X-ray image suffers from severe ambiguity of internal body information, where numbers of CT volumes can exactly match the same input X-rays once projected onto 2D. It seems to be unsolvable if we look for general solutions with traditional CT reconstruction algorithms. However, human body anatomy is well constrained and we may be able to learn the mapping from X-rays to CT from a large training set through machine learning technology, especially deep learning (DL) methods. Recently, the generative adversarial network (GAN) \cite{goodfellow2014generative} has been used for cross-modality image transfer in medical imaging \cite{bahrami2016convolutional, burgos2015robust, nie2017medical,zhang2018translating} and has demonstrated the effectiveness. However, the previous works only deal with the input and output data having the same dimension. Here we propose X2CT-GAN that can reconstruct CT from biplanar X-rays, surpassing the data limitations of different modalities and dimensionality (Fig. \ref{motivation}). 


The purpose of this work is not to replace CT with X-rays. Though the proposed method can reconstruct the general structure accurately, small anatomies still suffer from some artifacts. However, the proposed method may find some niche applications in clinical practice. For example, we can measure the size of major organs (\eg, lungs, heart, and liver) accurately, or diagnose ill-positioned organs on the reconstructed CT scan. It may also be used for dose planning in radiation therapy, or pre-operative planning and intra-operative guidance in minimally invasive intervention. It could be a nice enhancement of a low-cost X-ray machine as physicians may also get a CT-like 3D volume that has certain clinical values.

Though the proposed network can also be used to reconstruct CT from a single X-ray, we argue that using biplanar X-rays is a more practical solution.
First, CT reconstruction from a single X-ray subjects to too much ambiguity while biplanar X-rays offer additional information from both views that is complementary to each other. More accurate results, 4 dB improvement in peak signal-to-noise ratio (PSNR), are achieved in our comparison experiment.
Second, biplanar X-ray machines are already clinically available, which can capture two orthogonal X-ray images simultaneously. And, it is also clinically practicable to capture two orthogonal X-rays with a mono-planar machine, by rotating the X-ray apparatus to a new orientation for the second X-ray imaging. 

One practical issue to train X2CT-GAN is lacking of paired X-ray and CT \footnote{Sometimes X-rays are captured as topogram before a CT scan. However, without the calibrated back-projection matrix, we cannot perfectly align the two modalities.}.
It is expensive to collect such paired data from patients and it is also unethical to subject patients to additional radiation doses. In this work, we train the network with synthesized X-rays generated from large public-available chest CT datasets \cite{armato2011lung}. Given a CT volume, we simulate two X-rays, one from the posterior-anterior (PA) view and the other from the lateral view, using the digitally reconstructed radiographs (DRR) technology \cite{milickovic2000ct}. Although DRR synthesized X-rays are quite photo-realistic, there still exits a gap between real and synthesized X-rays, especially in finer anatomy structures, \eg, blood vessels. Therefore we further resort CycleGAN \cite{zhu2017unpaired} to learn the genuine X-ray style that can be transferred to the synthesized data. More information about the style transfer operation can be found in supplement materials. 

\begin{figure}
    \centering
    \begin{center}
       \includegraphics[width=1.0\linewidth]{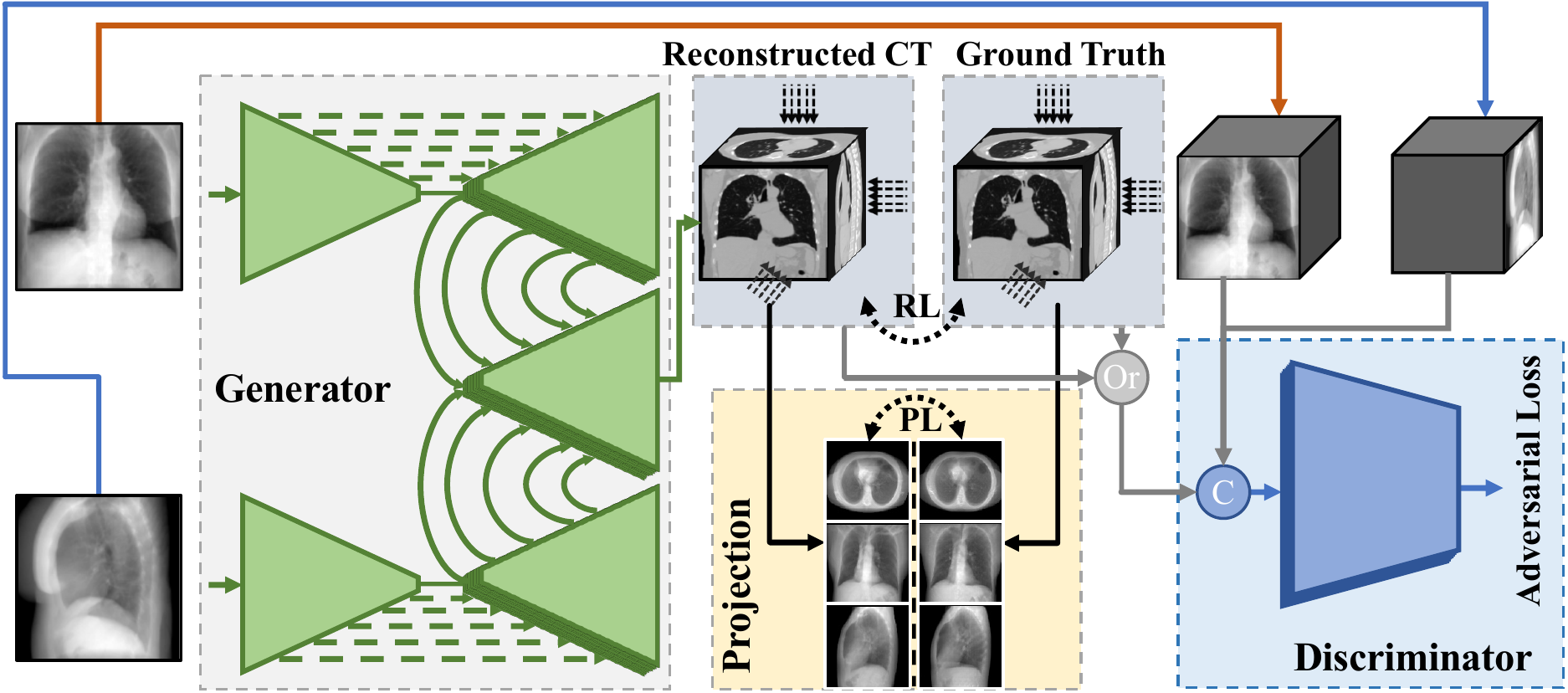}
    \end{center}
       \caption{Overview of the X2CT-GAN model. RL and PL are abbreviations of the reconstruction loss and projection loss.}
    \label{model}
\end{figure}

To summarize, we make the following contributions:
\begin{itemize}
    \item We are the first to explore CT reconstruction from biplanar X-rays with deep learning. To fully utilize the input information from two different views, a novel feature fusion method is proposed.
    \vspace{-1ex}
		
	  \item We propose X2CT-GAN, as illustrated in Fig. \ref{model}, to increase the data dimension from input to output (\ie, 2D X-ray to 3D CT), which is not addressed in previous research on GAN.
    \vspace{-1ex}
				
    \item We propose a novel skip connection module that could bridge 2D and 3D feature maps more naturally.
    \vspace{-1ex}		
		
    \item We use synthesized X-rays to learn the mapping from 2D to 3D, and CycleGAN to transfer real X-rays to the synthesized style before feeding into the network. Therefore, although our network is trained with synthesized X-rays, it can still reconstruct CT from real X-rays.
    \vspace{-1ex}
				
    \item Compared to other reconstruction algorithms using visible light~\cite{choy20163d, fan2017point, jiang2018gal}, our X-ray based approach can reconstruct both surface and internal structures.
\end{itemize}

\begin{figure*}[t]
    \centering
       \includegraphics[width=1.0\linewidth]{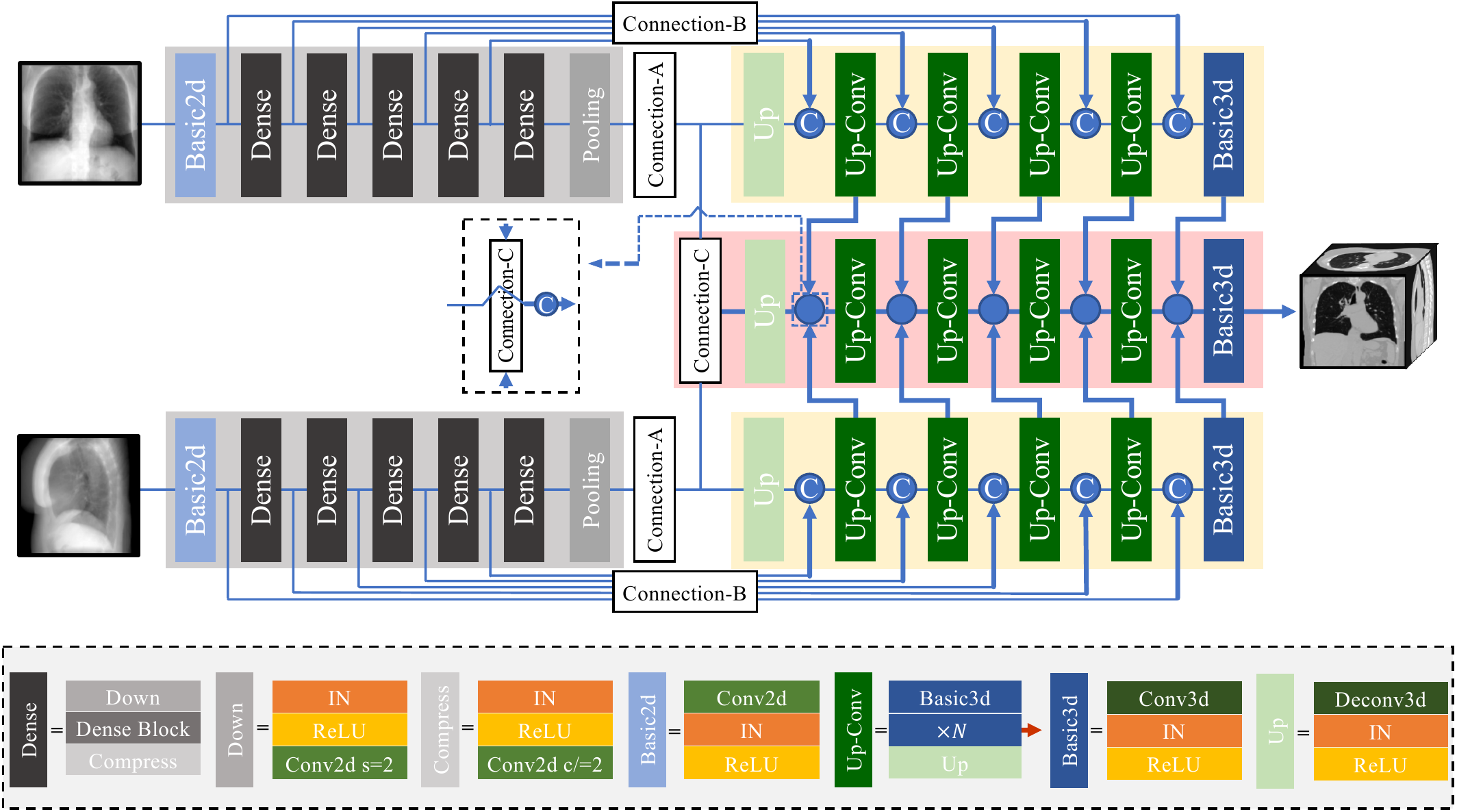}
       \caption{Network architecture of the X2CT-GAN generator. Two encoder-decoder networks with the same architecture work in parallel for posterior-anterior (PA) and lateral X-rays, respectively. Another fusion network between these two encoder-decoder networks is responsible for fusing information coming from two views. For more details about Connection-A, B and C, please refer to Fig. \ref{connect}.}
    \label{generator}
\end{figure*}

\section{Related Work}
\textbf{Cross-Modality Transfer} 
A DL based model often suffers from lacking enough training data so as to fall into a suboptimal point during training or even overfit the small dataset. To alleviate this problem, synthetic data has been used to boost the training process \cite{shrivastava2017learning, zhang2018translating}. So synthesizing realistic images close to the target distribution is a critic premise. Previous research such as pix2pix \cite{isola2017image} could do the pixel level image to image transfer and CycleGAN \cite{zhu2017unpaired} has the ability to learn the mapping between two unpaired datasets. 
In medical imaging community, quite some efforts have been put into this area to transfer a source modality to a target modality, \eg, 3T MRI to 7T MRI \cite{bahrami2016convolutional}, MRI to CT \cite{burgos2015robust, nie2017medical}, MRI and CT bidirectional transfer \cite{zhang2018translating} \etc. 
Our approach differs from the previous cross-modality transfer works in two ways.
First, in all the above works, the dimensions of the input and output are consistent, \eg, 2D to 2D or 3D to 3D.
Here, we want to transfer 2D X-rays to a 3D volume.
To handle this challenge, we propose X2CT-GAN, which incorporates two mechanisms to increase the data dimension.
Second, our goal is to reconstruct accurate 3D anatomy from biplanar X-rays with clinical values instead of enriching the training set.
A photo-realistic image (\eg, one generated from pure noise input~\cite{goodfellow2014generative}) may already be beneficial for training.
However, our application further requires the image to be anatomically accurate as well.

\textbf{3D Model Extraction from 2D Projections}
3D model extraction from 2D projections is a well studied topic in computer vision \cite{choy20163d, fan2017point, jiang2018gal}.
Since most objects are opaque to light, only the outer surface model can be reconstructed.
X-ray can penetrate most objects (except thick metal) and different structures overlay on a 2D image.
Therefore, the methods used in 3D model extraction from X-rays are quite different to those used in the computer vision community.
Early in 1990, Caponetti and Fanelli reconstructed a bone model from two X-rays based on back-lighting projections, polygon mesh and B-spline interpolation \cite{caponetti19903d}. 
In recent years, several works have investigated the reconstruction of bones, rib cages and lungs through statistical shape models or other prior knowledge \cite{dworzak20103d, aubert20163d, karade20153d, lamecker2006atlas, koehler20103, koehler20113d, koehler2010knowledge}.
Different to ours, these methods could not generate a 3D CT-like image.
Furthermore, although they may be able to get precise models, if we generalize these to reconstruct other organs, an elaborate geometric model has to be prepared in advance, which limits their application scenarios.

\textbf{CT Reconstruction from X-ray}
Classical CT reconstruction algorithms, \eg, filtered back projection and iterative reconstruction methods~\cite{Herman2009CT}, require hundreds of X-rays captured during a full rotational scan of the body.
Methods based on deep learning have also been used to improve the performance in recent works \cite{wurfl2016deep, hammernik2017deep}. 
The input of \cite{wurfl2016deep} is an X-ray sinogram, while ours are human readable biplanar X-rays. 
And, \cite{hammernik2017deep} mainly deals with the limited-angle CT compensation problem. 
More relevant to our work is \cite{henzler2017single}, which uses a convolutional neural network (CNN) to predict the underlying 3D object as a volume from a single-image tomography.
However, we argue that a single X-ray is not enough to accurately reconstruct 3D anatomy since it is subject to too much ambiguity.
For example, we can stretch or flip an object along the projection direction without changing the projected image.
As shown in our experiments, biplanar X-rays with two orthogonal projections can significantly improve the reconstruction accuracy, benefiting from more constraints provided by an additional view.
Furthermore, the images reconstructed by \cite{henzler2017single} are quite blurry, thus with limited clinical values. Combining adversarial training and reconstruction constraints, our method could extract much finer anatomical structures (\eg blood vessels inside lungs), which significantly improves the visual quality.

\section{Objective Functions of X2CT-GAN}
GAN \cite{goodfellow2014generative} is a recent  proposal to effectively train a generative model that has demonstrated the ability to capture real data distribution. Conditional GAN \cite{mirza2014conditional}, as an extension of the original GAN, further improves the data generation process by conditioning the generative model on additional inputs, which could be class labels, partial data, or even data from a different modality. Inspired by the successes of conditional GANs, we propose a novel solution to train a generative model that can reconstruct a 3D CT volume from biplanar 2D X-rays. In this section, we first introduce several loss functions that are used to constrain the generative model. 
\vspace{1pt}
\subsection{Adversarial Loss}
The original intention of GAN is to learn deep generative models while avoiding approximating many intractable probabilistic computations that arise in other strategies, \ie, maximum likelihood estimation. The learning procedure is a two-player game, where a discriminator $D$ and a generator $G$ would compete with each other. The ultimate goal is to learn a generator distribution $p_G(x)$ that matches the real data distribution $p_{data}(x)$. An ideal generator could generate samples that are indistinguishable from the real samples by the discriminator.
More formally, the minmax game is summarized by the following expression:
\begin{equation}
    \label{raw_GAN}
    \begin{aligned}
    \min\limits_{G}\max\limits_{D}V(G,D) = &\mathbb{E}_{{x}\sim{p_{data}}}[\log{D({x})}]+\\&\mathbb{E}_{{z}\sim{noise}}[\log{(1-D(G({z})))}],
    \end{aligned}
\end{equation}
where $z$ is sampled from a noise distribution. 

As we want to learn a non-linear mapping from X-rays to CT, the generated CT volume should be consistent with the semantic information provided by the input X-rays. After trying different mutants of the conditional GAN, we find out that LSGAN \cite{mao2017least} is more suitable for our task and apply it to guide the training process. The conditional LSGAN loss is defined as:
\begin{equation}
    \label{LSGAN_D}
    \begin{aligned}
    \mathcal{L}_{LSGAN}(D) = \frac{1}{2}[&\mathbb{E}_{y\sim{p(CT)}}(D(y|x)-1)^2+\\&\mathbb{E}_{x\sim{p(Xray)}}(D(G(x)|x)-0)^2],\\
    \mathcal{L}_{LSGAN}(G) = \frac{1}{2}[&\mathbb{E}_{x\sim{p(Xray)}}(D(G(x)|x)-1)^2],
    \end{aligned}
\end{equation}

where $x$ is composed of two orthogonal biplanar X-rays, and $y$ is the corresponding CT volume. Compared to the original objective function defined in Eq. (\ref{raw_GAN}), LSGAN replaces the logarithmic loss with a least-square loss, which helps to stabilize the adversarial training process and achieve more realistic details. 

\subsection{Reconstruction Loss}
The conditional adversarial loss tries to make prediction look real. However, it does not guarantee that $G$ can generate a sample maintaining the structural consistency with the input. Moreover, CT scans, different from natural images that have more diversity in color and shape, require higher precision of internal structures in 3D. Consequently, an additional constraint is required to enforce the reconstructed CT to be voxel-wise close to the ground truth. Some previous work has combined the reconstruction loss \cite{pathak2016context} with the adversarial loss and got positive improvements. We also follow this strategy and acquire a high PSNR as shown in Table \ref{quant}.  Our reconstruction loss is defined as MSE:
\begin{equation}
    \label{re_loss}
    \begin{aligned}
    \mathcal{L}_{re} = 
                      \mathbb{E}_{x,y}\lVert{y-G(x)}\rVert_{2}^2.
    \end{aligned}
\end{equation}

\subsection{Projection Loss}
The aforementioned reconstruction loss is a voxel-wise loss that enforces the structural consistency in the 3D space. To improve the training efficiency, more simple shape priors could be utilized as auxiliary regularizations. Inspired by \cite{jiang2018gal}, we impel 2D projections of the predicted volume to match the ones from corresponding ground-truth in different views. Orthogonal projections, instead of perspective projections, are carried out to simplify the process as this auxiliary loss focuses only on the general shape consistency, not the X-ray veracity. We choose three orthogonal projection planes (axial,  coronal, and sagittal, as shown in Fig. \ref{model}, following the convention in the medical imaging community). Finally, the proposed projection loss is defined as below: 

\begin{equation}
    \label{pl_Loss}
    \begin{aligned}
    \mathcal{L}_{pl} = \frac{1}{3}[&\mathbb{E}_{x,y}\lVert{P_{ax}(y)-P_{ax}(G(x))}\rVert_{1}+\\
                            &\mathbb{E}_{x,y}\lVert{P_{co}(y)-P_{co}(G(x))}\rVert_{1}+\\
                            &\mathbb{E}_{x,y}\lVert{p_{sa}(y)-P_{sa}(G(x))}\rVert_{1}],
    \end{aligned}
\end{equation}
where the $P_{ax}$, $P_{co}$ and $P_{sa}$ represent the projection in the axial, coronal, and sagittal plane, respectively. The $L1$ distance is used to enforce sharper image boundaries. 

\subsection{Total Objective}
Given the definitions of the adversarial loss, reconstruction loss, and projection loss, our final objective function is formulated as:
\begin{equation}
    \label{total_loss}
    \begin{aligned}
    &D^{*} = arg\min\limits_{D}\lambda_{1}\mathcal{L}_{LSGAN}(D),\\
    &G^{*} = arg\min\limits_{G}[\lambda_{1}\mathcal{L}_{LSGAN}(G)+\lambda_{2}\mathcal{L}_{re}+\lambda_{3}\mathcal{L}_{pl}],
    \end{aligned}
\end{equation}
where $\lambda_{1}$, $\lambda_{2}$ and $\lambda_{3}$ control the relative importance of different loss terms. In our X-ray to CT reconstruction task, the adversarial loss plays an important role of encouraging local realism of the synthesized output, but global shape consistency should be prioritized during the optimization process. Taking this into consideration, we set $\lambda_{1}=0.1, \lambda_{2}=\lambda_{3}=10$ in our experiments. 

\section{Network Architecture of X2CT-GAN}
In this section, we introduce our proposed network designs that are used in the 3D CT reconstruction task from 2D biplanar X-rays. Similar to other 3D GAN architectures, our method involves a 3D generator and a 3D discriminator. These two models are alternatively trained with the supervision defined in previous section.  

\subsection{Generator}
The proposed 3D generator, as illustrated in Fig. \ref{generator}, consists of three individual components: two encoder-decoder networks with the same architecture working in parallel for posterior-anterior (PA) and lateral X-rays respectively, and a fusion network. The encoder-decoder network aims to learn the mapping from the input 2D X-ray to the target 3D CT in the feature space, and the fusion network is responsible for reconstructing the 3D CT volume with the fused biplanar information from the two encoder-decoder networks. Since the training process in our reconstruction task involves circulating information between input and output from two different modalities and dimensionalities, several modifications of the network architecture are made to adapt to the challenge.   

\textbf{Densely Connected Encoder} Dense connectivity \cite{huang2017densely} has a compelling advantage in the feature extraction process. To optimally utilize information from 2D X-ray images, we embed dense modules to generator's encoding path. As shown in Fig. \ref{generator}, each dense module consists of a down-sampling block (2D convolution with stride=2), a densely connected convolution block and a compressing block (output channels halved). The cascaded dense modules encode different level information of the input image and pass it to the decoder along different shortcut paths. 

\begin{figure}
	\centering
	\begin{center}
		\includegraphics[width=0.8\linewidth]{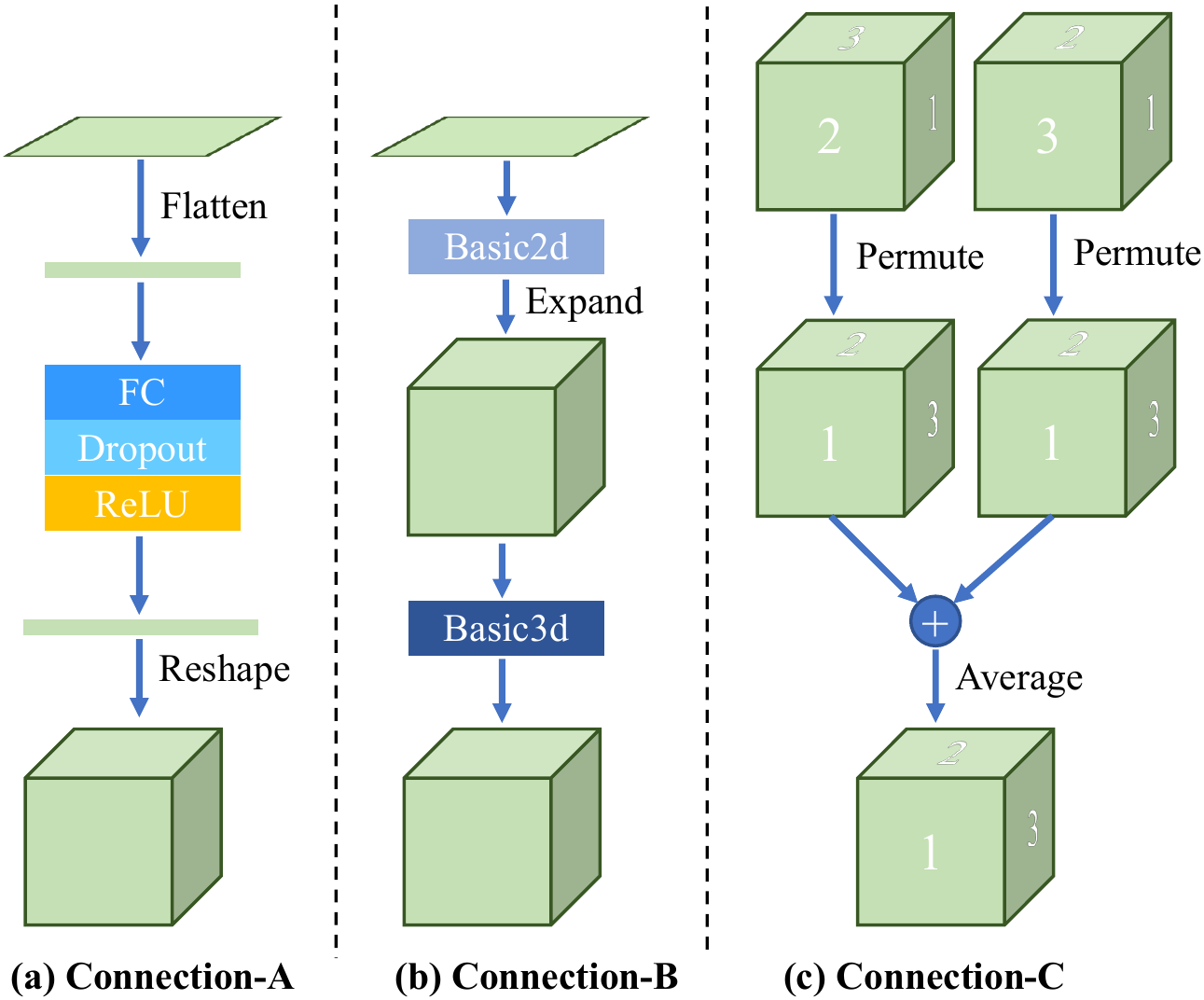}
	\end{center}
	\caption{Different types of connections. Connection-A and Connection-B aim to increase dimensionality of feature maps, and Connection-C is for fusing information from two different views.}
	\label{connect}
\end{figure}

\textbf{Bridging 2D Encoder and 3D Decoder} 
Some existing encoder-decoder networks \cite{isola2017image, ledig2017photo} link encoder and decoder by means of convolution. There is no obstacle in a pure 2D or 3D encode-decode process, but our special 2D to 3D mapping procedure requires a new design to bridge the information from two dimensionalities. Motivated by \cite{zhu2018image}, we extend fully connected layer to a new connection module, named Connection-A (Fig. \ref{connect}a), to bridge the 2D encoder and 3D decoder in the middle of our generator. To better utilize skip connections in the 2D-3D generator, we design another novel connection module, named Connection-B (Fig. \ref{connect}b), to shuttle low-level features from encoder to decoder.

To be more specific, Connection-A achieves the 2D-3D conversion through fully-connected layers, where the last encoder layer's output is flattened and elongated to a 1D vector that is further reshaped to 3D. However, most of the 2D spatial information gets lost during such conversion so that we only use Connection-A to link the last encoder layer and first decoder layer. For the rest of skip connections, we use Connection-B and take following steps: 1) enforce the channel number of the encoder being equal to the one on the corresponding decoder side by a basic 2D convolution block; 2) expand the 2D feature map to a pseudo-3D one by duplicating the 2D information along the third axis; 3) use a basic 3D convolution block to encode the pseudo-3D feature map. The abundant low-level information shuttled across two parts of the network imposes strong correlations on the shape and appearance between input and output.    

\textbf{Feature Fusion of Biplanar X-rays}
 As a common sense, a 2D photograph from frontal view could not retain lateral information of the object and vice versa. In our task, we resort biplanar X-rays captured from two orthogonal directions, where the complementary information could help the generative model achieve more accurate results. Two encoder-decoder networks in parallel extract features from each view while the third decoder network is set to fuse the extracted information and output the reconstructed volume. As we assume the biplanar X-rays are captured within a negligible time interval, meaning no data shift caused by patient motions, we can directly average the extracted features after transforming them into the same coordinate space, as shown in Fig. \ref{connect}c. Any structural inconsistency between two decoders' outputs will be captured by the fusion network and back-propagated to two networks. 

\subsection{Discriminator}
 PatchGANs have been used frequently in recent works \cite{li2016precomputed, isola2017image, ledig2017photo, zhu2017unpaired, wang2017high} due to the good generalization property. We adopt a similar architecture in our discriminator network from Phillip \etal \cite{isola2017image}, named as 3DPatchDiscriminator. It consists of three $conv3d-norm-relu$ modules with $stride=2$ and $kernel size=4$, another $conv3d-norm-relu$ module with $stride=1$ and $kernel size=4$, and a final $conv3d$ layer. Here, $conv3d$ denotes a 3D convolution layer; $norm$ stands for an instance normalization layer \cite{instancenorm}; and $relu$ represents a rectified linear unit \cite{glorot2011deep}. The proposed discriminator architecture improves the discriminative capacity inherited from the PatchGAN framework and can distinguish real or fake 3D volumes.  

\subsection{Training and Inference Details}
The generator and discriminator are trained alternatively following the standard process \cite{goodfellow2014generative}. We use the Adam solver \cite{kingma2014adam} to train our networks. The initial learning rate of Adam is 2$e$-4, momentum parameters $\beta_{1}=0.5$ and $\beta_{2}=0.99$. After training 50 epochs, we adopt a linear learning rate decay policy to decrease the learning rate to 0. We train our model for a total of 100 epochs.

As instance normalization \cite{instancenorm} has been demonstrated to be superior to batch normalization \cite{ioffe2015batch} in image generation tasks, we use instance normalization to regularize intermediate feature maps of our generator. At inference time, we observe that better generating results could be obtained if we use the statistics of the test batch itself instead of the running average of training batches, as suggested in \cite{isola2017image}.
Constrained by GPU memory limit, the batch size is set to one in all our experiments.

\section{Experiments}
    In this section, we introduce an augmented dataset built on LIDC-IDRI \cite{armato2011lung}. We evaluate the proposed X2CT-GAN model with several widely used metrics, \eg, peak signal-to-noise ratio (PSNR) and structural similarity (SSIM) index. To demonstrate the effectiveness of our method, we reproduce a baseline model named 2DCNN \cite{henzler2017single}. Fair comparisons and comprehensive analysis are given to demonstrate the improvement of our proposed method over the baseline and other mutants. Finally, we show the real-world X-ray evaluation results of X2CT-GAN. Input images to X2CT-GAN are resized to $128 \times 128$ pixels, while the input of 2DCNN is $256 \times 256$ pixels as suggested by \cite{henzler2017single}. The output of all models is set to $128 \times 128 \times 128$ voxels.

\begin{figure}[t]
    \centering
       \includegraphics[width=1.0\linewidth]{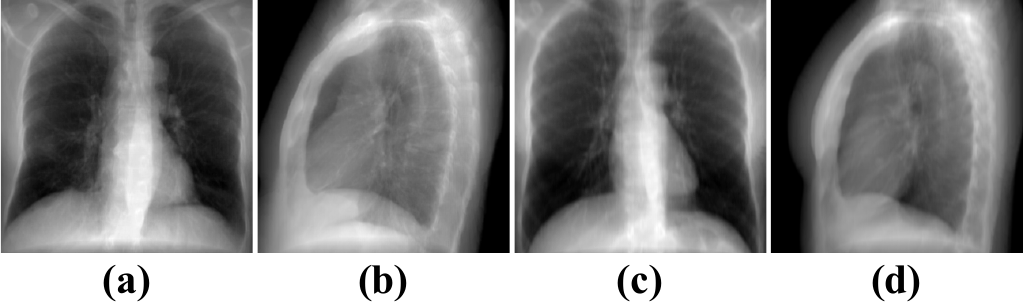}
       \caption{DRR \cite{milickovic2000ct} simulated X-rays. (a) and (c) are simulated PA view X-rays of two subjects, (b) and (d) are the corresponding lateral views.}
    \label{drr}
\end{figure}
    
\begin{figure*}[t]
	\centering
	\includegraphics[width=0.9\linewidth]{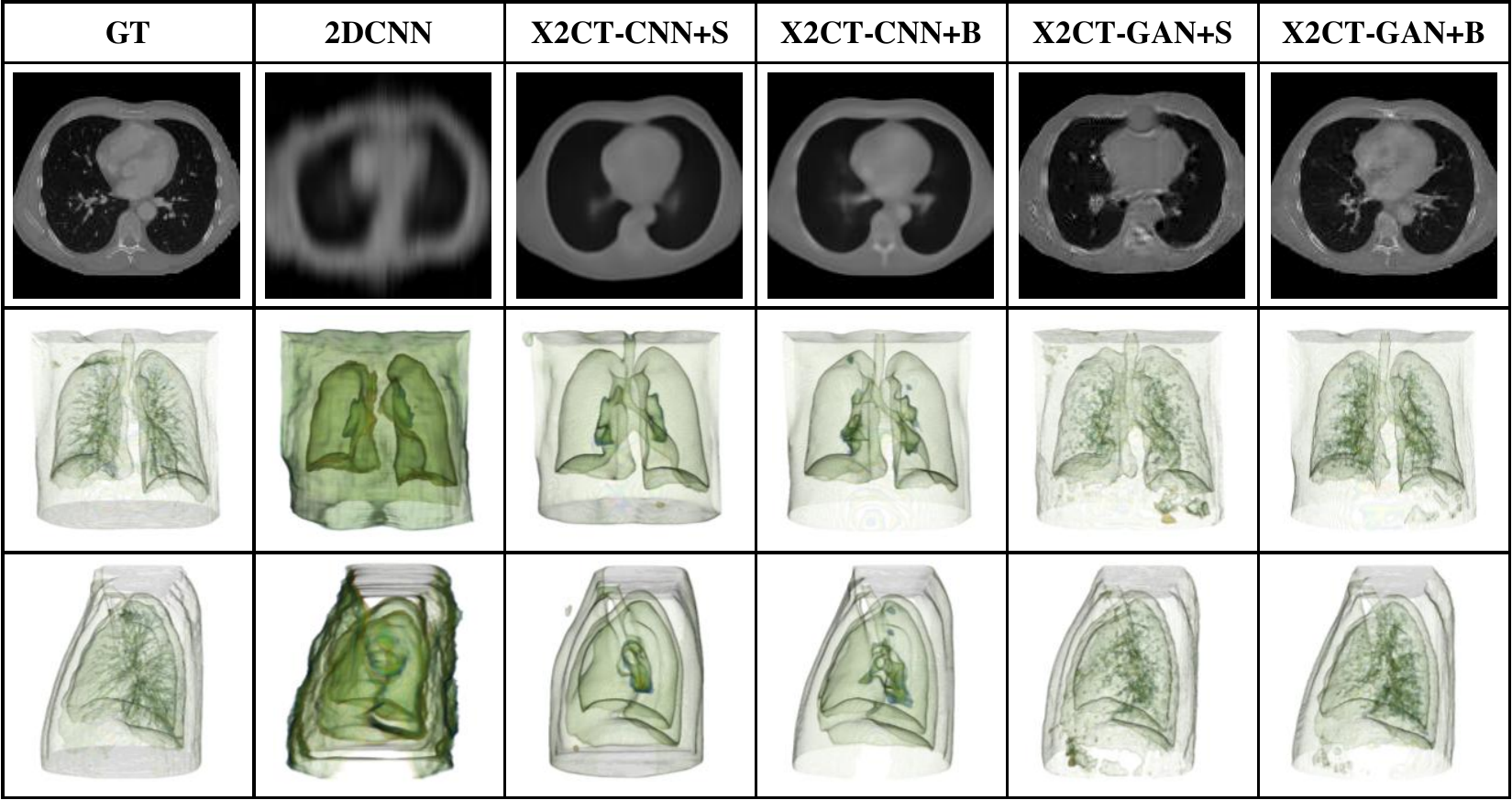}
	\caption{Reconstructed CT scans from different approaches. 2DCNN is our reproduced baseline model \cite{henzler2017single}; X2CT-CNN is our generator network optimized by the MSE loss alone and X2CT-GAN is our GAN-based model optimized by total objective. `+S' means single-view X-ray input and `+B' means biplanar X-rays input. The first row demonstrates axial slices generated by different models. The last two rows are 3D renderings of generated CT scans in the PA and lateral view, respectively.}
	\label{quali}
\end{figure*}

\subsection{Datasets}
\textbf{CT and X-ray Paired Dataset} 
Ideally, to train and validate the proposed CT reconstruction approach, we need a large dataset with paired X-rays and corresponding CT reconstructions.
Furthermore, the X-ray machine needs to be calibrated to get an accurate projection matrix.
However, no such dataset is available and it is very costly to collect such real paired dataset.
Therefore, we take a real CT volume and use the digitally reconstructed radiographs (DRR) technology \cite{milickovic2000ct} to synthesize corresponding X-rays, as shown in Fig. \ref{drr}. 
It is much cheaper to collect such synthesized datasets to train our networks.
To be specific, we use the publicly available LIDC-IDRI dataset \cite{armato2011lung}, which contains 1,018 chest CT scans.
The heterogeneous of imaging protocols result in different capture ranges and resolutions. 
For example, the number of slices varies a lot for different volumes. The resolution inside a slice is isotropic but also varies for different volumes. All these factors lead to a nontrivial reconstruction task from 2D X-rays. To simplify, we first resample the CT scans to the $1 \times 1 \times 1$ mm$^3$ resolution.
Then, a $320 \times 320 \times 320$ mm$^3$ cubic area is cropped from each CT scan. 
We randomly select 916 CT scans for training and the rest 102 CT scans are used for testing.

\textbf{Mapping from Real to Synthetic X-rays} 
Although DRR synthetic X-rays are quite photo-realistic, there is still a gap between the real and synthetic X-rays, especially for those subtle anatomical structures, \eg, blood vessels.
Since our networks are trained with synthesized X-rays, a sub-optimal result will be obtained if we directly feed a real X-ray into the network.
We propose to perform style transfer to map real X-rays to the synthesized style.
Without paired dataset of real and synthesized X-rays, we exploit CycleGAN \cite{zhu2017unpaired} to learn the mapping.
We collected 200 real X-rays and randomly selected 200 synthetic X-rays from the training set of the paired LIDC dataset.

\subsection{Metrics}
\textbf{PSNR} is often used to measure the quality of reconstructed digital signals \cite{oriani2011qpsnr}. 
Conventionally, CT value is recorded with 12 bits, representing a range of [0, 4095] (the actual Hounsfield unit equals the CT value minus 1024) \cite{bankman2008handbook}, which makes PSNR an ideal criterion for image quality evaluation. 

\textbf{SSIM} is a metric to measure the similarity of two images, including brightness, contrast and structure \cite{wang2004image}. Compared to PSNR, SSIM can match human's subjective evaluation better.

\subsection{Qualitative Results}
\begin{figure}[t]
    \centering
       \includegraphics[width=1.0\linewidth]{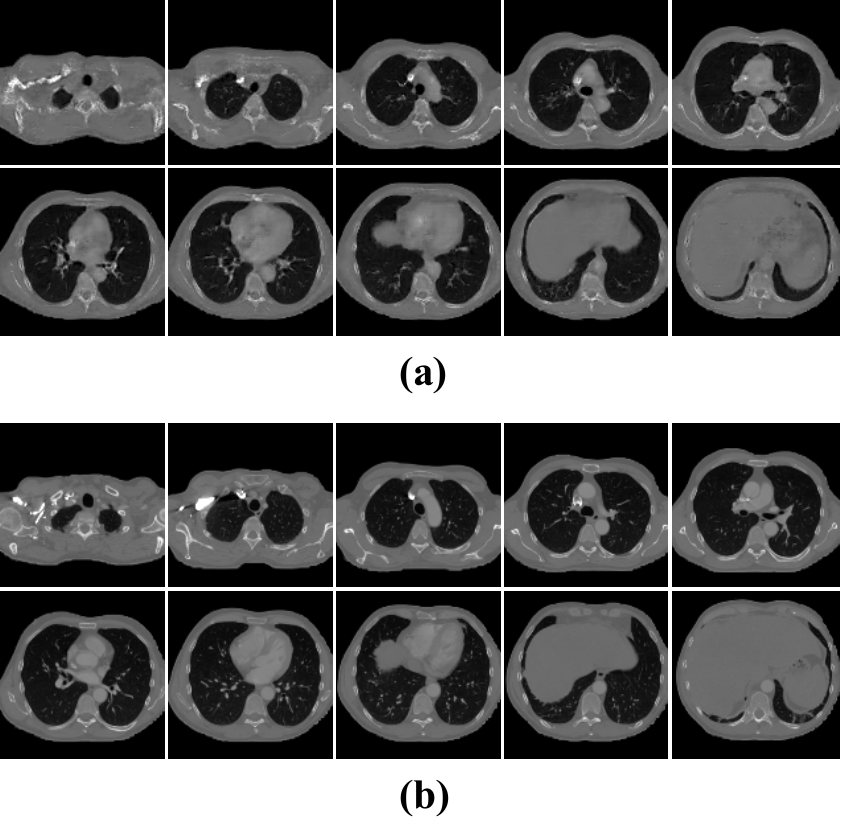}
       \caption{Examples of reconstructed CT slices (a) and corresponding groundtruth (b). As could be seen, our method reconstructs the shape and appearance of major anatomical structures accurately.}
    \label{ctslices}
\end{figure}

We first qualitatively evaluate CT reconstruction results shown in Fig. \ref{quali}, where X2CT-CNN is the proposed network supervised only by the reconstruction loss while X2CT-GAN is the one trained with full objectives; `+S' means single-view X-ray input and `+B' means biplanar X-rays input.
For comparison, we also reproduce the method proposed in \cite{henzler2017single} (referred as 2DCNN in Fig. \ref{quali}) as the baseline, one of very few published works tackling the X-ray to CT reconstruction problem using deep learning. Since 2DCNN is designed to deal with single X-ray input, no biplanar results are shown.
From the visual quality evaluation, it is obvious to see the differences. First of all, 2DCNN and X2CT-CNN generate very blurry volumes while X2CT-GAN maintains small anatomical structures. Secondly, though missing reconstruction details, X2CT-CNN+S generates sharper boundaries of large organs (\eg, heart, lungs and chest wall) than 2DCNN. Last but not least, models trained with biplanar X-rays outperform the ones trained with single view X-ray. More reconstructed CT slices could be found in Fig. \ref{ctslices}.

\subsection{Quantitative Results}
Quantitative results are summarized in Table \ref{quant}.
Biplanar inputs significantly improve the reconstruction accuracy, about 4 dB improvement for both X2CT-CNN and X2CT-GAN, compared to single X-ray input.
It is well known that the GAN models often sacrifice MSE-based metrics to achieve visually better results.
This phenomenon is also observed here.
However, by tuning the relative weights of the voxel-level MSE loss and semantic-level adversarial loss is our cost function, we can make a reasonable trade-off.
For example, there is only 1.1 dB decrease in PSNR from X2CT-CNN+B to X2CT-GAN+B, while the visual image quality is dramatically improved as shown in Fig. \ref{quali}.
We argue that visual image quality is as important as (if not more important than) PSNR in CT reconstruction since eventually the images will be read visually by a physician.

\begin{table}
\caption{Quantitative results. 2DCNN is our reproduced model from \cite{henzler2017single}; X2CT-CNN is our generator network optimized by the MSE loss alone; and X2CT-GAN is our GAN-based model optimized by total objective. `+S' means single-view X-ray input and `+B' means biplanar X-rays input.}
\begin{center}
\begin{tabular}{|c|c|c|c|}
\hline
Methods & PSNR (dB) & SSIM \\ \hline\hline
2DCNN   & 23.10($\pm$0.21) & 0.461($\pm$0.005)  \\ \hline
X2CT-CNN+S    & 23.12($\pm$0.02) & 0.587($\pm$0.001)  \\ \hline
X2CT-CNN+B    & \textbf{27.29($\pm$0.04)} & \textbf{0.721($\pm$0.001)}  \\ \hline
X2CT-GAN+S    & 22.30($\pm$0.10) & 0.525($\pm$0.004)  \\ \hline
X2CT-GAN+B    & \textbf{26.19($\pm$0.13)} & \textbf{0.656($\pm$0.008)}  \\ \hline
\end{tabular}
\end{center}
\label{quant}
\end{table}

\subsection{Ablation Study}
\textbf{Analysis of Proposed Connection Modules}
To validate the effectiveness of proposed connection modules, we also perform an ablation study in the setting of X2CT-CNN. As shown in Table \ref{connection_abl}, single view input with Connection-B achieves 0.7 dB improvement in PSNR. The biplanar input, even without skip connections, surpasses the single view due to the complementary information injected to the network. And in our biplanar model, Connection-B and Connection-C are interdependent so that we regard them as one module. As can be seen, the biplanar model with this module surpasses other combinations by a large margin both in PSNR and SSIM.

\begin{table}
\small
\caption{Evaluation of different connection modules. `XC' denotes X2CT-CNN model without the proposed Connection-B and Connection-C module. `+S' means the model's input is a single-view X-ray and `+B' means biplanar X-rays. `CB' and `CC' denote Connection-B and Connection-C respectively as shown in Fig. \ref{connect}.}
\begin{center}
\begin{tabular}{|c|c|c|c|c|c|}
\hline
\multicolumn{4}{|c|}{Combination} & \multicolumn{2}{c|}{Metrics} \\ \hline\hline
    XC+S       & XC+B       & CB         & CC         & PSNR(dB)          & SSIM \\ \hline
    \checkmark &            &            &            & 22.46($\pm$0.02) & 0.549($\pm$0.002)\\ 
    \checkmark &            & \checkmark &            & 23.12($\pm$0.02) & 0.587($\pm$0.001)\\ 
               & \checkmark &            &            & 24.84($\pm$0.05) & 0.620($\pm$0.003)\\ 
               & \checkmark & \checkmark & \checkmark & \textbf{27.29($\pm$0.04)} & \textbf{0.721($\pm$0.001)}\\
\hline
\end{tabular}
\end{center}
\label{connection_abl}
\end{table}

\textbf{Different Settings in GAN Framework}
The effects of different settings in the GAN framework are summarized in Table \ref{gan_abl}. As the first row shows, adversarial loss alone performs poorly on PSNR and SSIM due to the lack of strong constraints. The most significant improvement comes from the reconstruction loss being added to the GAN framework. Projection loss and the conditional information bring additional improvement slightly.

\begin{table}
\small
\caption{Evaluation of different settings in the GAN framework. `RL' and `PL' denote the reconstruction and projection loss, respectively. `CD' means that input X-ray information is fed to the discriminator to achieve a conditional GAN.}
\begin{center}
\begin{tabular}{|c|c|c|c|c|c|}
\hline
\multicolumn{4}{|c|}{Formulation} & \multicolumn{2}{c|}{Metrics} \\ \hline\hline
    GAN        & RL         & PL         & CD         & PSNR(dB)         & SSIM     \\ \hline
    \checkmark &            &            &            & 17.38($\pm$0.36)&
0.347($\pm$0.022)\\ 
    \checkmark & \checkmark &            &            & 25.82($\pm$0.08)& 0.645($\pm$0.001)\\ 
    \checkmark & \checkmark & \checkmark &            & 26.05($\pm$0.02)& 0.645($\pm$0.002)\\ 
    \checkmark & \checkmark & \checkmark & \checkmark & \textbf{26.19($\pm$0.13)}& \textbf{0.656($\pm$0.008)}\\
\hline
\end{tabular}
\end{center}
\label{gan_abl}
\end{table}

\begin{figure}[t]
    \centering
       \includegraphics[width=1.0\linewidth]{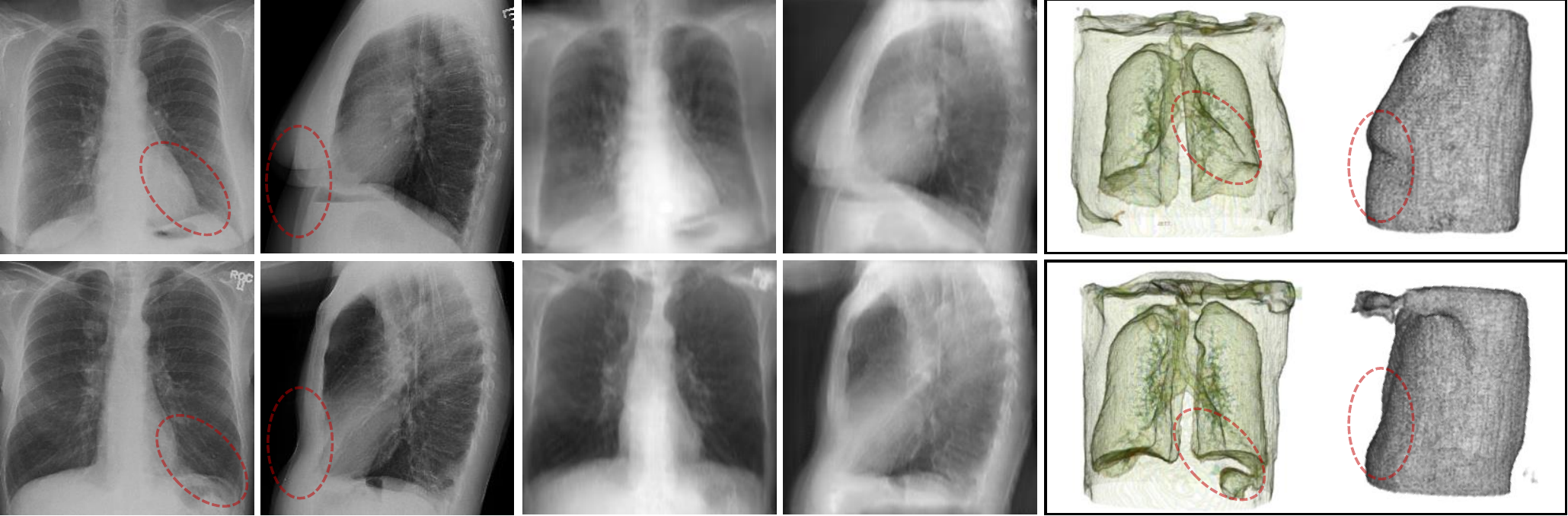}
       \caption{CT reconstruction from real-world X-rays. Two subjects are shown here. The first and second columns are real X-rays in two views. The following two columns are transformed X-rays by CycleGAN~\cite{zhu2017unpaired}. The last two columns show 3D renderings of reconstructed internal structures and surfaces. Dotted ellipses highlight regions of high-quality anatomical reconstruction.}
    \label{realresult}
\end{figure}

\subsection{Real-World Data Evaluation}
Since the ultimate goal is to reconstruct a CT scan from real X-rays, we finally evaluate our model on real-world data, despite the model is trained on synthetic data. 
As we have no corresponding 3D CT volumes for real X-rays, only qualitative evaluation is conducted. 
Visual results are presented in Fig. \ref{realresult}, we could see that the reconstructed lung and surface structures are quite plausible. 

\section{Conclusions}
In this paper, we explored the possibility of reconstructing a 3D CT scan from biplanar 2D X-rays in an end-to-end fashion. 
To solve this challenging task, we combined the reconstruction loss, the projection loss and the adversarial loss in the GAN framework. 
Moreover, a specially designed generator network is exploited to increase the data dimension from 2D to 3D. Our experiments qualitatively and quantitatively demonstrate that biplanar X-rays are superior to single view X-ray in the 3D reconstruction process.
For future work, we will collaborate physicians to evaluate the clinical value of the reconstructed CT scans, including measuring the size of major organs and dose planning in radiation therapy, etc.

{\small
\bibliographystyle{ieee}
\bibliography{egbib}
}

\clearpage

\section{Appendix}

\begin{figure*}
 	\centering
	\includegraphics[width=1.0\linewidth]{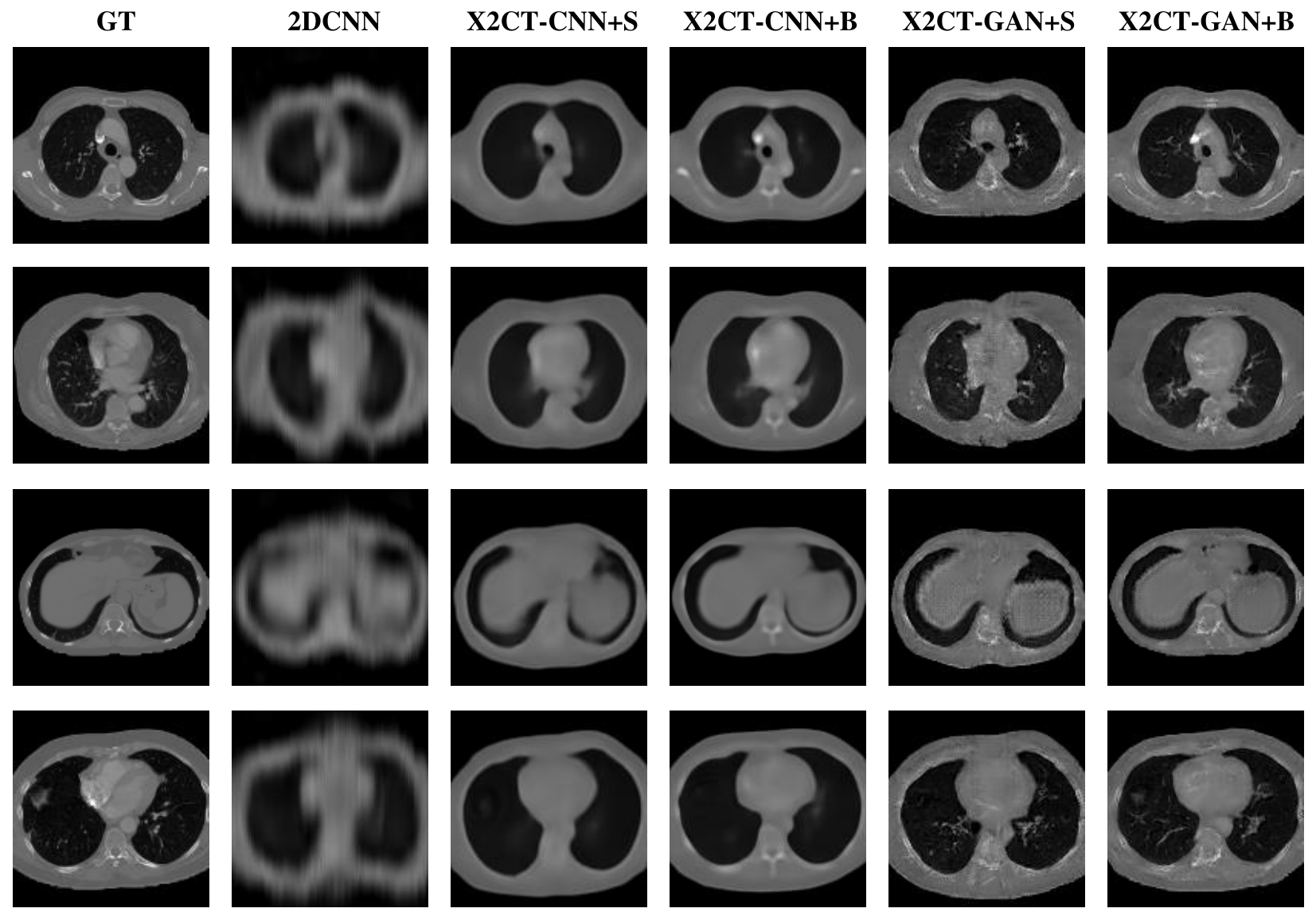}
	\caption{CT slices sampled from four subjects. The first column presents the groundtruths, and the corresponding reconstruction results achieved by different approaches are in rows.}
	\label{sp_slices1}
\end{figure*}

\subsection{Introduction}
\begin{figure*}
    \centering
   \includegraphics[width=1.0\linewidth]{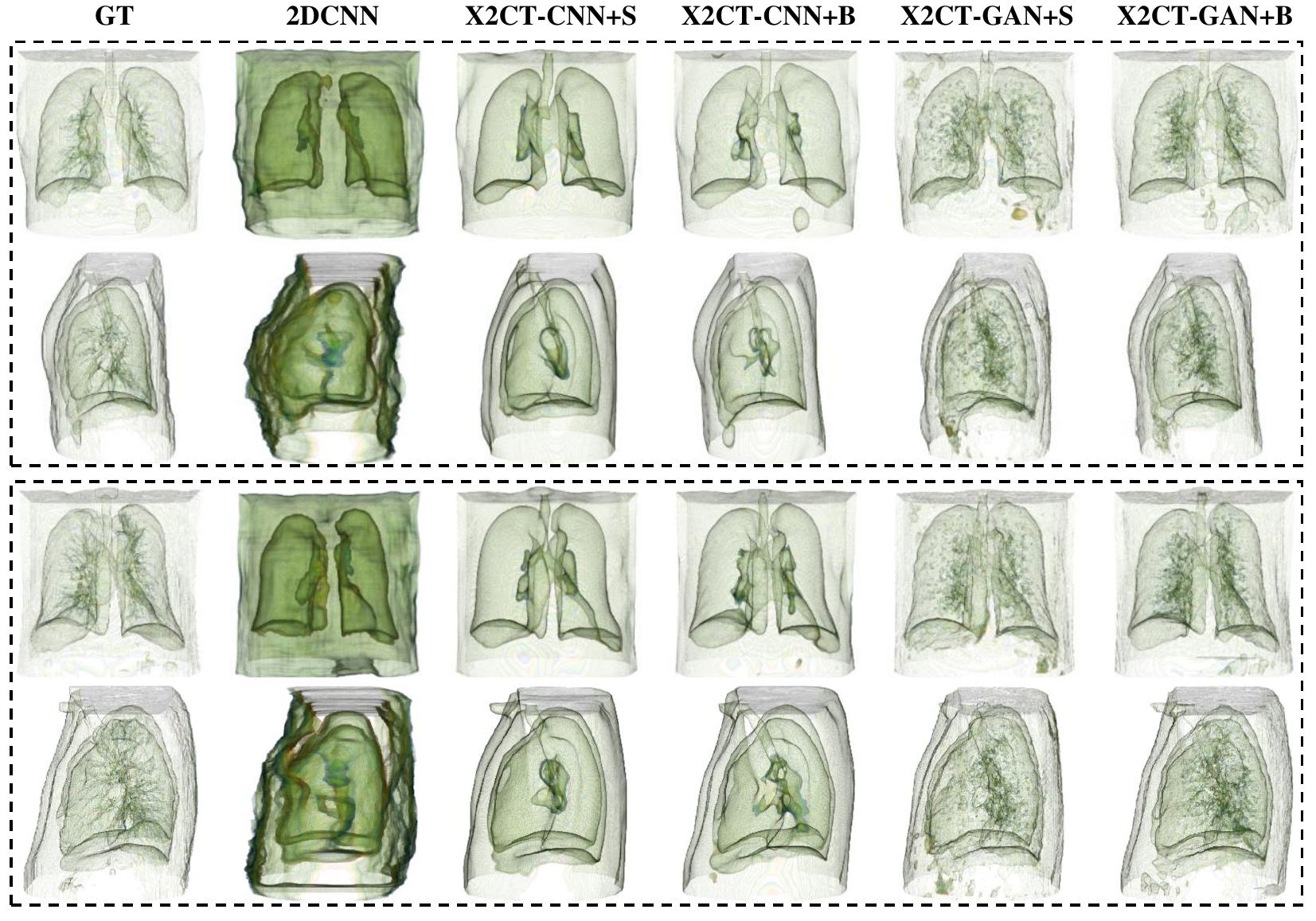}
   \caption{Another two cases about the 3D renderings of CT scans reconstructed by different approaches. Similarly, PA view and lateral view are shown here. As can be seen, models with biplanar input have more accurate results viewed in lateral. And GAN-based models extract much finer anatomical structures (e.g. blood vessels inside lungs).}
   \label{3Drenderings}
\end{figure*}

In this supplementary material, more visual results are provided. In Fig. \ref{sp_slices1}, we present four CT slices sampled from four subjects and their corresponding reconstruction results achieved by different approaches. In Fig. \ref{3Drenderings}, we present additional 3D renderings of two subjects.

\begin{figure*}
    \centering
   \includegraphics[width=1.0\linewidth]{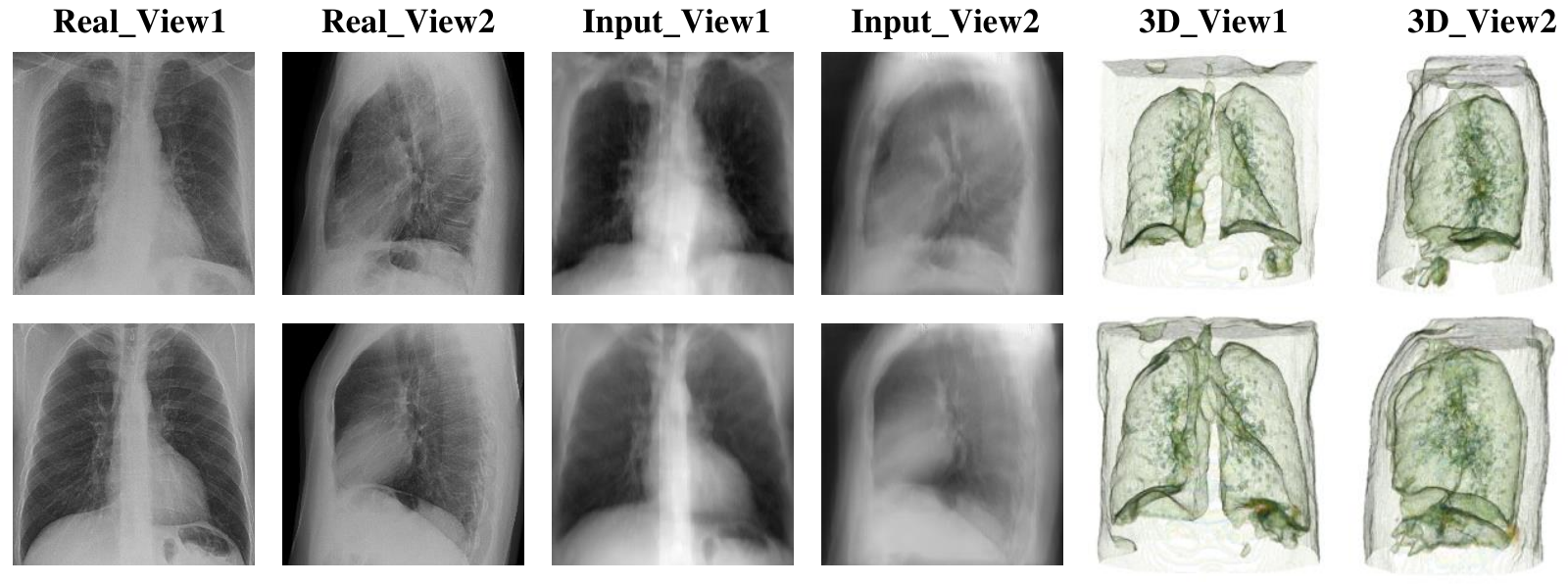}
   \caption{Another two cases about CT reconstruction from real-world X-rays. The first and second columns are real X-rays in two views. The following two columns are transformed X-rays by CycleGAN \cite{zhu2017unpaired}, and they are the inputs of our X2CT-GAN model. The last two columns show 3D renderings of reconstructed CT scan in PA view and lateral view.}
   \label{sp_realEval}
\end{figure*}

\subsection{Style Transfer Between Real X-rays and Synthetic X-rays}
Although DRR \cite{milickovic2000ct} synthesized X-rays are quite photo-realistic, there still exits a gap between real and synthetic X-rays, especially in finer anatomy structures, \eg, blood vessels. Therefore we further resort CycleGAN \cite{zhu2017unpaired} to learn the genuine X-ray style that can be transferred to the synthetic data. To achieve this, we collected 200 real X-rays and randomly selected 200 synthetic X-rays from the training set of the paired LIDC dataset. The network architecture and parameters initialization are kept same as in \cite{zhu2017unpaired}. We train the CycleGAN model for a total of 200 epochs. 
In Fig. \ref{sp_realEval}, another two testing results of real-world X-rays are given. 

\end{document}